  \documentclass[review, number, sort&compress, 12pt]{elsarticle}

	\usepackage{url}

 \usepackage{graphicx}

\usepackage{amssymb,amsbsy,amsmath}

\usepackage{textcomp}
\usepackage{upgreek}

\journal{NIM A}

\begin{document}

\begin{frontmatter}



\title{Position-sensitive detection of ultracold neutrons with an imaging camera and its implications to spectroscopy}

\author[LANL]{Wanchun Wei}
\author[LANL]{L. J. Broussard}
\author[LANL]{M. A. Hoffbauer}
\author[LANL]{M. Makela}
\author[LANL]{C. L. Morris}%
\author[LANL]{Z. Tang}
\author[IU]{E. R. Adamek}
\author[IU]{N. B. Callahan}
\author[LANL]{S. M. Clayton}
\author[NCSU]{C. Cude-Woods}
\author[LANL]{S. Currie}
\author[NCSU]{E. B. Dees}
\author[VA]{X. Ding}
\author[ILL]{P. Geltenbort}
\author[Caltech]{K. P. Hickerson}
\author[TTU]{A. T. Holley}
\author[LANL]{T. M. Ito}
\author[NCSU]{K. K. Leung}
\author[IU]{C.-Y. Liu}
\author[LANL]{D. J. Morley}
\author[LANL]{Jose D. Ortiz}
\author[LANL]{R. W. Pattie, Jr.}
\author[LANL]{J. C. Ramsey}
\author[LANL]{A. Saunders}
\author[LANL]{S. J. Seestrom}
\author[Russia]{E. I. Sharapov}
\author[LANL]{S. K. Sjue}
\author[NCSU]{J. Wexler}
\author[LANL]{T. L. Womack}
\author[NCSU]{A. R. Young}
\author[NCSU]{B. A. Zeck}
\author[LANL]{Zhehui Wang\corref{cor1}}
\cortext[cor1]{Corresponding author: Z. Wang, email: zwang@lanl.gov}
\address[LANL]{Los Alamos National Laboratory, Los Alamos, NM 87545, USA}
\address[IU]{Indiana University, Bloomington, IN 47405, USA}
\address[NCSU]{North Carolina State University, Raleigh, NC 27695, USA}
\address[VA]{Virginia Polytechnic Institute and State University, Blacksburg, VA 24061, USA}
\address[ILL]{Institut Laue Langevin, 38042 Grenoble, France}
\address[Caltech]{California Institute of Technology, Pasadena, CA 91125, USA}
\address[TTU]{Tennessee Technological University, Cookeville, TN 38505, USA}
\address[Russia]{Joint Institute for Nuclear Research, 141980, Dubna, Russia}

\begin{abstract}
Position-sensitive detection of ultracold neutrons (UCNs) is demonstrated using an imaging charge-coupled device (CCD) camera.  A spatial resolution less than 15 $\mu$m has been achieved, which is equivalent to an UCN energy resolution below 2 pico-electron-volts through the relation $\delta E = m_0g \delta x$. Here, the symbols $\delta E$, $\delta x$, $m_0$ and $g$ are the energy resolution, the spatial resolution, the neutron rest mass and the gravitational acceleration, respectively. \newpage \noindent  A multilayer surface convertor described previously is used to capture UCNs and then emits visible light for CCD imaging.  Particle identification and noise rejection are discussed through the use of light intensity profile analysis. This method allows different types of UCN spectroscopy and other applications.
\end{abstract}

\pagebreak[4]


\end{frontmatter}



\newpage
\section{Introduction}
\label{sec:1}
A multilayer surface detector for ultra-cold neutrons (UCNs),  or neutrons with kinetic energies below 340 nano-electron-volts (neV), has recently been reported~\cite{Wang:2015}. The multilayer detector front end consists of a boron-10 ($^{10}$B) layer on the top of a ZnS:Ag phosphor layer. The $^{10}$B layer, up to several hundred nanometers thick, is exposed to vacuum and directly captures UCNs. A phosphor layer several microns thick is sufficient to stop the charged ions from the $^{10}$B(n,$\alpha$)$^7$Li neutron capture reaction, while thin enough so that light due to $\alpha$ and $^7$Li escapes for detection by photomultiplier tubes (PMTs). 

We extend the previous work by using an imaging charge-coupled device (CCD) camera instead of a photo-multiplier tube (PMT) to detect UCN-induced visible light which peaks around 450 nm. We show that the position resolution of UCN detection is comparable to the size of individual pixels or 15 $\mu$m. The multilayer construction of the detector front end remains the same as before. The use of optics allows the separation of the neutron converter in vacuum from the camera in air. Applications of this method include UCN spectroscopy and UCN-based materials research. For spectroscopy, a resolution of 15-$\mu$m corresponds to an energy resolution of less than 2 pico-electron-volts (peV) for some designs (see Sec.~\ref{sec:spec} below). Because of the availability of imaging cameras with smaller pixels for scientific and consumer use, further improvements are possible that would yield a spatial resolution of few microns, or sub-peV energy resolutions. It should be mentioned that at a resolution of a few microns, which are comparable to the stopping distances of MeV $\alpha$'s and $^7$Li, the details of ion stopping in ZnS phosphor may no longer be ignored. Examples of fine spatial resolution for ultra-cold, cold and thermal neutrons can be found in different context~\cite{Lehmann:2004,Hussey:2005,Kawasaki:2010,Williams:2012,He:2013,Jenke:2011,Blostein:2015}. Another interesting development is the adoption of a webcam for UCN detection by removing the cover glass on the sensor array~\cite{Lauer:2011}. Besides CCD and complementary metal-oxide-semiconductor (CMOS) imaging sensors, hybrid pixelated silicon devices with an application specific integrated circuit such as TimePix were also reported for position-sensitive UCN detection and spectroscopy~\cite{Jakubek:2009}. 

Below, we first present the experimental setup, followed by experimental data and analysis, and finally a discussion of UCN spectroscopy and materials applications. 

\section{Experimental setup}
\begin{figure}[thbp] 
   \centering
   \includegraphics[width=4in]{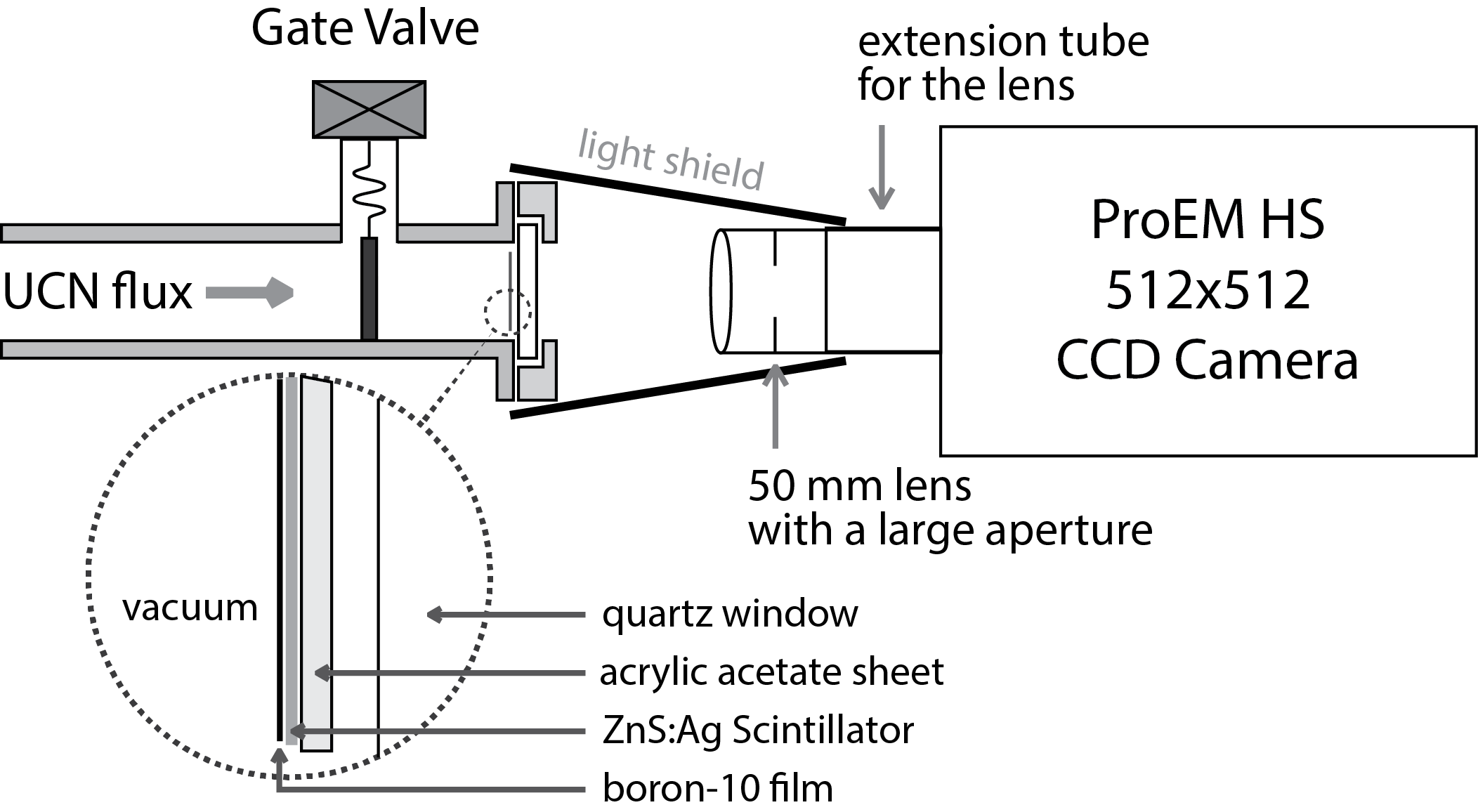}
  \caption{Schematic of the experimental setup for position-senstive measurement of UCNs. The UCN flux is regulated by a gate valve (GV) before reaching the multilayer UCN detector front end, which sits against the quartz vacuum window. The multilayer UCN detector front end consists of a thin layer of boron-10 on top of a layer of ZnS:Ag scintillator on top of an acrylic plastic sheet, as shown in magnified view. A Nikon Nikkor 50 mm $f$/1.2 lens was used to focus the scintillation light onto the CCD camera.}
   \label{fig1:sch}
 \end{figure}

The experiment was carried out in the UCN beamline in the Los Alamos Neutron Science Center (LANSCE)~\cite{Saunders:2013}. As shown in Figure~\ref{fig1:sch}, the neutron-to-photon converter film was placed against the quartz window at the end of the test port. An $\sim$80 nm $^{10}$B film faced the incoming flux of UCNs, which were controlled by a gate valve (GV). The products of UCNs captured on the $^{10}$B film were primarily $^7$Li at 0.84 MeV and $\alpha$ at 1.47 MeV. They induced blue scintillation light in the silver-doped ZnS film. The light peaked at 450 nm with a characteristic decay time of about 200 ns. The ion stopping ranges of $^7$Li and $\alpha$ are calculated to be within a few microns in both the $^{10}$B and ZnS:Ag films. If fully stopped in the scintillator, each $\alpha$ particle at 1.47 MeV is estimated to yield 7.4$\times$10$^4$ photons~\cite{Wang:2015}. The scintillation light is first transmitted out of the transparent acrylic plastic substrate through a small gap, {\it i.e.}, small to human eye but large relative to the wavelengths of scintillation light, passed through the quartz window, and then partially collected by a 50-mm camera lens before reaching an electron-multiplying CCD (EMCCD) imaging sensor.

 The camera (ProEM-HS made by Princeton Instruments) has a back-illuminated EMCCD sensor. The full sensor array has 512$\times$512 mono-chrome pixels with 16-bit well depth. Each pixel is 16 $\mu$m$\times$16 $\mu$m in physical size. A Nikon Nikkor 50 mm lens has a maximum $f$-number ($f$) of $f$/1.2 and a 46-mm wide clear aperture. Coupling the lens with a stack of extension tubes enables close photography of the film (within a few cm). A length calibration showed that each pixel corresponds to 15 $\mu$m at the film location, giving a full field-of-view of 7.7$\times$7.7 mm$^2$. The camera has a built-in thermoelectric cooling system, which reduces the sensor noise level greatly when the sensor temperature is below -70 $^0$C. During the measurement, the optical window and camera lens are enclosed in an ambient light shield (two layers of black Tedlar sheets), while the camera body remains in open air for ventilation. 

The number of photons that can be ultimately recorded by the camera is estimated as follows. The indices of refraction of ZnS and acrylic plastic are about 2.5 and 1.5 respectively. Since the indices of refraction decrease monotonically from ZnS to acrylic plastic to the vacuum gap, total internal reflections will occur at each interface. The critical angle that allows the light to transmit out of the converter film is about 23.6 degrees, or $\theta_c$ = asin(1/2.5), at the interface of ZnS and acrylic. The associated solid angle fraction is 4.1\%, which is also the fraction of the total scintillation light that can come out of the converter film for isotropic emission. The light then passes through the quartz vacuum window, where it experiences no solid angle loss but a transmission loss of about 10\% at 450 nm due to absorption and reflection. Hence, 3.7\% of the scintillation light can exit the window and spread out in a 2$\pi$ solid angle. The camera lens is positioned about 31 mm away from the window. The solid angles to the lens from points in the 7.7$\times$7.7 mm$^2$ image area vary slightly around 0.4$\pi$, corresponding to the optical setup having about 20\% of the light collection efficiency. In the experiment, the lens aperture was set at $f$/2.0 to balance the depth-of-field (for ease of focusing) and the field-of-view. The $f$/2.0 aperture only allows 40\% of the light collection that would otherwise be obtained at $f$/1.2. The QE of the CCD sensor is about 85\% at 450 nm. Taking all the loss factors into account, an upper limit of about 190 photons generated by an $\alpha$ particle of 1.47 MeV, or about 0.2\% of the 7.4$\times$10$^4$ photons, can be detected by the image sensor.

\section{Results and Data Analysis}
\label{sec:DC}
To determine our spatial resolution, we analyze the images of individual scintillation events, modelled as a 2D Gaussian response function.  

\begin{figure}[htbp] 
 \centering
 \includegraphics[width=6in,angle=0]{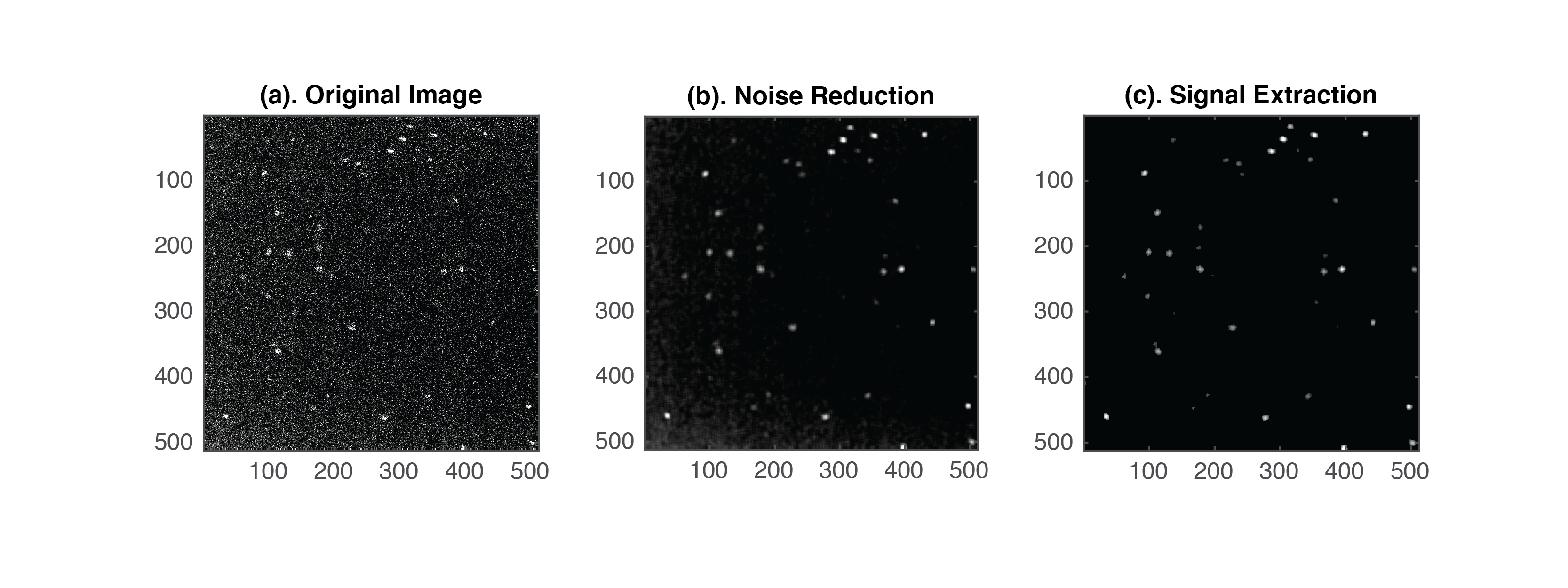}
   \caption{An example of a raw image and processing. (a.) The raw image. (b.) The processed image after initial noise reduction by averaging over the neighboring 7$\times$7 pixels. (c.) The processed image after signal extraction and background removal.}
  \label{fig2:img1}
\end{figure}

An example raw image and its processed versions are shown in Figure~\ref{fig2:img1}. The image was taken with an exposure time of 1200 ms. The raw image shows a poor signal-to-noise ratio due to background noise. An initial noise reduction is achieved by averaging over the 7$\times$7 neighboring pixels. As shown in Figure~\ref{fig2:img1}b, the sharpness of the image is sacrificed to improve the signal-to-noise ratio. 
The image is evenly divided into 8$\times$8 sub-regions where the mean background noise level is relatively flat. Only pixels above a fixed threshold are kept, and the mean background of the subregion is subtracted (Figure~\ref{fig2:img1}c).  The threshold is set to a few times the standard deviation above the mean pixel value of the image. Thus we generally separate signal pixels from those of the background. In order to avoid double counting of signals on the boundary of the sub-regions, the pixel extraction is repeated on the final image. 
Limitations of this approach include (a.) false identification of some uneven backgrounds as signals, and (b.) failure in separating overlapping signals if two neutron captures are too close to each other (pulse pile-up). 

\begin{figure}[htbp] 
 \centering
 \includegraphics[width=4in,angle=0]{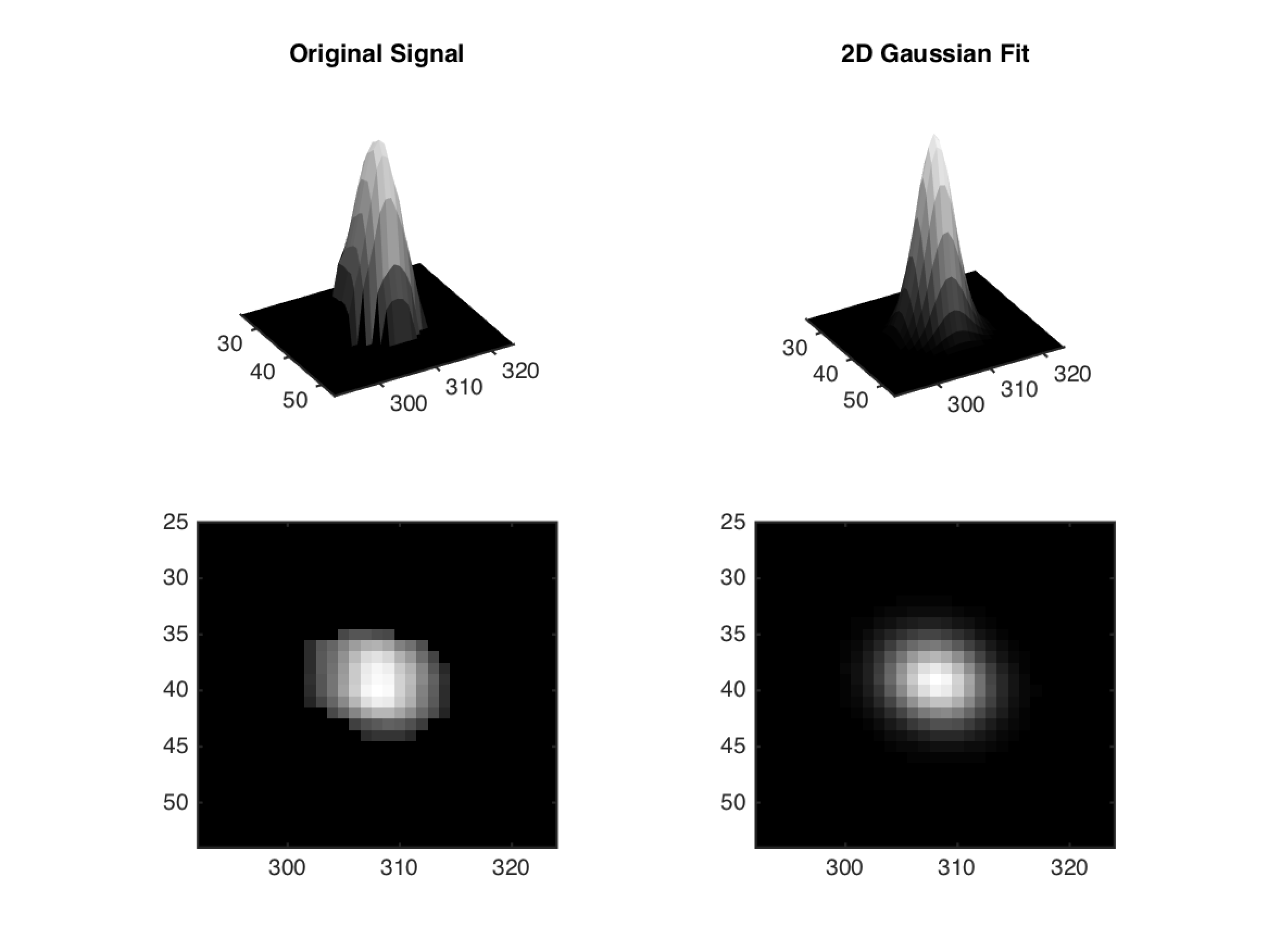}
   \caption{An example of the two-dimensional (2D) Gaussian fitting of the scintillation signal from an UCN capture. The pre-processed original signal, as described in Figure~\ref{fig2:img1} is in the left column (top and bottom for the 3D and 2D views) and the 2D Gaussian fit is on the right.}
  \label{fig3:img2}
\end{figure}

Scintillator light is emitted along the paths of MeV $\alpha$ and $^7$Li ions in the scintillator, which have ion-stopping ranges of a few microns. The 15-$\mu$m pixel resolution in this work is at least a few times the ion ranges, and therefore can not resolve the details of ion stopping. Furthermore, only certain neutron events are at sharp focus due to the shallow depth of focus. To account for the different image blurring mechanisms, we use a 2D elliptical Gaussian function to fit each spot. The signal $S$ is centered around ($x_0$, $y_0$), the position of neutron capture,

\begin{equation}
S = A \exp\left[-a(x-x_0)^2 + 2b(x-x_0)(y-y_0) - c(y-y_0)^2\right]+B,
\end{equation}
where

\begin{equation}
\begin{aligned}
a&=\frac{\cos^2 \theta}{2\sigma_1^2} + \frac{\sin^2 \theta}{2\sigma_2^2}, \\
b&=- \frac{\sin 2\theta}{ 4\sigma_1^2}+\frac{\sin 2\theta}{ 4\sigma_2^2},\\
c&=\frac{\sin^2 \theta}{2\sigma_1^2} + \frac{\cos^2 \theta}{2\sigma_2^2}.
\end{aligned}
\label{eq:gauss2}
\end{equation}
Here $A$ is the height of the Gaussian blob, $B$ is the value of the base plane, $\sigma_1$ and $\sigma_2$ are the widths along the major and minor semi-axes, and $\theta$ is the angle between the major semi-axis and the x-axis in the Cartesian coordinate system. A typical fit is compared with the original signal as shown in Figure~\ref{fig3:img2}, where ($x_0$, $y_0$) = (308.13, 39.14) pixels, $A$ = 37.6, $\sigma_1$ = 3.15 pixels, and $\sigma_2$ = 2.51 pixels.
 The signal intensity is taken from the integral over the processed signal, which is assumed to be proportional to the number of photons detected. For instance, the integrated intensity from the data in Figure~\ref{fig3:img2} is 1811.6. 

\begin{figure}[htbp] 
 \centering
 \includegraphics[width=4in,angle=0]{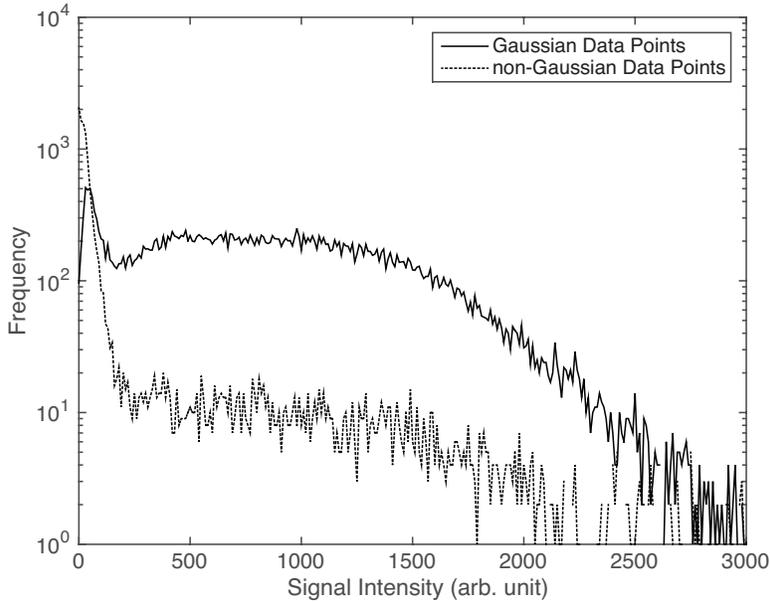} 
   \caption{Pulse height spectra for Gaussian and non-Gaussian signals. Only Gaussian signals are counted as real neutron-induced signals.} 
  \label{fig4}
\end{figure}

We have taken data at different exposure times ranging from 30 ms up to a few seconds. Since our algorithm for signal identification works best for a low signal density, data at exposure times longer than 1200 ms are not included in the following analysis. There are a total of 4.5$\times$10$^4$ events identified by the analysis. The quality of the 2D Gaussian fit is used to exclude events due to uneven background, dark currents and piled-up signals. The latter two are rare events for exposure times under 1200 ms. About 76\% of these preliminary identified events are well fitted by the 2D Gaussian. Pulse height spectra of peak intensities for both Gaussian and non-Gaussian events are shown in Figure~\ref{fig4}. 
The quality of the fit removes a large fraction of background events (low-intensity) and a small fraction of real UCN events (from pile-up rejection).  Signals with poor fit are excluded from the analysis that follows.

 To investigate the nature of background noise, we compare data taken for a fixed 800-ms exposure time with and without a UCN flux. We note that, although the UCN flux was turned off to the detector, 
 the spallation source was still running. A total of 112 out of the 300 frames were taken without a UCN flux. 
 
\begin{figure}[htbp] 
 \centering
 \includegraphics[width=4in,angle=0]{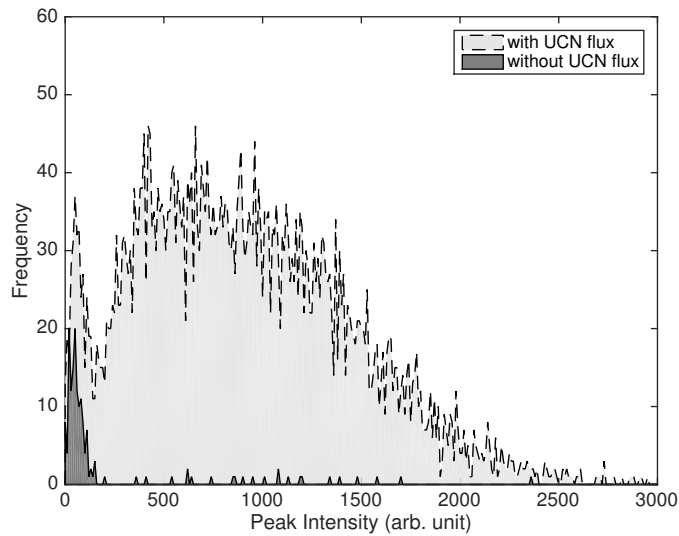} 
   \caption{Pulse height spectra of the integrated peak intensities for the GV-open periods (light-shaded) and for the GV-closed periods (dark-shaded). The same data set as shown in Figure~\ref{fig5} below is used. It is clear that the noise dominates the pulses below 180.}
  \label{fig6}
\end{figure}

 In Figure~\ref{fig6}, we plot the pulse height spectra of integrated signal intensity in the GV-open and GV-closed cases, respectively. The spectrum looks qualitatively similar to those obtained previously by PMT-based detectors, as expected for a camera sensor array with a uniform response to light. 
The GV-closed dataset is dominated by background, and is used to set a lower intensity threshold. 

\begin{figure}[htbp] 
 \centering
 \includegraphics[width=4in,angle=0]{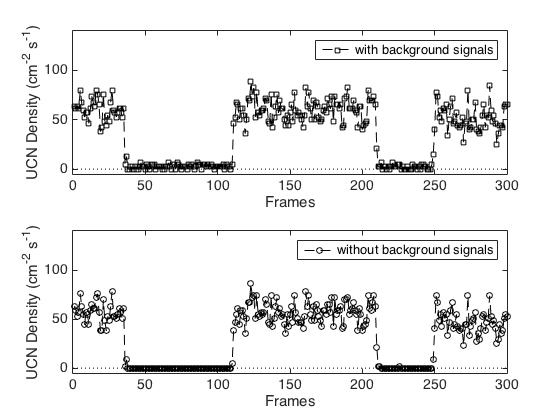} 
   \caption{UCN flux density as a function of frame (equivalent to time) using the movie mode of the camera. The exposure time is fixed at 800 ms for each image frame. The upper plot includes all signals well-fit by a 2D Gaussian.  The lower plot imposes an lower intensity threshold.} 
  \label{fig5}
\end{figure}

The UCN density per frame is plotted in Figure~\ref{fig5}. The two GV-closed periods are clearly identifiable with essentially no UCN counts. 

\begin{figure}[htbp] 
 \centering
 \includegraphics[width=4in,angle=0]{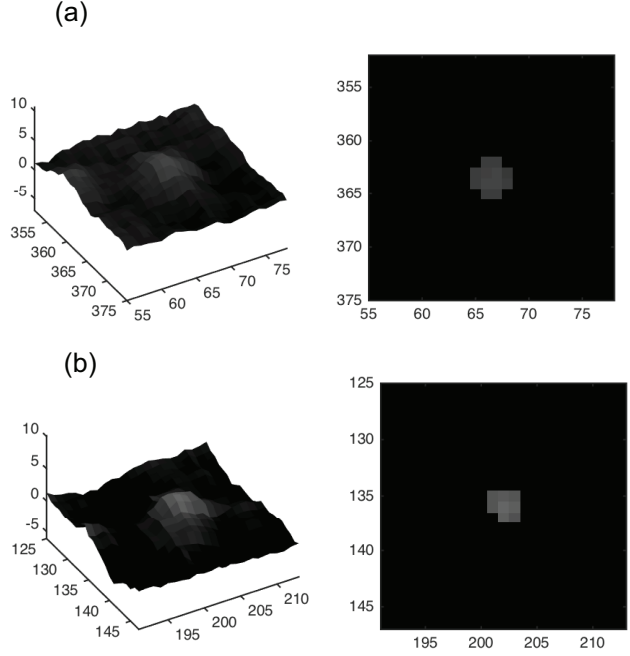} 
   \caption{Two examples of background noise (small amplitude less than 180) that are taken during the GV closed periods. (a.) The first row, a background noise with a Gaussian profile. (b.) The second row, a background noise with a non-Gaussian profile.}
  \label{fig7}
\end{figure}

In Figure~\ref{fig7}, two examples of images taken during GV-closed periods are presented: Figure~\ref{fig7}a shows a signal with a 2D Gaussian profile, while Figure~\ref{fig7}b shows a non-Gaussian profile. Both signals are non-UCN induced events on the image sensor since the GV was closed. In the GV-closed periods, the mean background rate is about 2.09 $\pm$ 0.17 events cm$^{-2}$ s$^{-1}$. In the GV-open period, 7\% of the total signals are below the threshold, with a mean rate of 4.07 $\pm$ 0.24 events cm$^{-2}$ s$^{-1}$
Up to 3.4\% of the total UCN signals may be lost using a threshold of 180 to exclude low intensity data. 
The mean UCN rate above the threshold is 53.89 $\pm$ 0.89 UCNs cm$^{-2}$ s$^{-1}$ in the GV-open periods, and 0.11 $\pm$ 0.05 UCNs cm$^{-2}$ s$^{-1}$ in the GV-closed periods. A PMT-based detector nearby that used a similar converter film read about 60 UCNs cm$^{-2}$ s$^{-1}$. 

\section{Discussion}
\subsection{Position resolution}
\begin{figure}[htbp] 
 \centering
 \includegraphics[width=4in,angle=0]{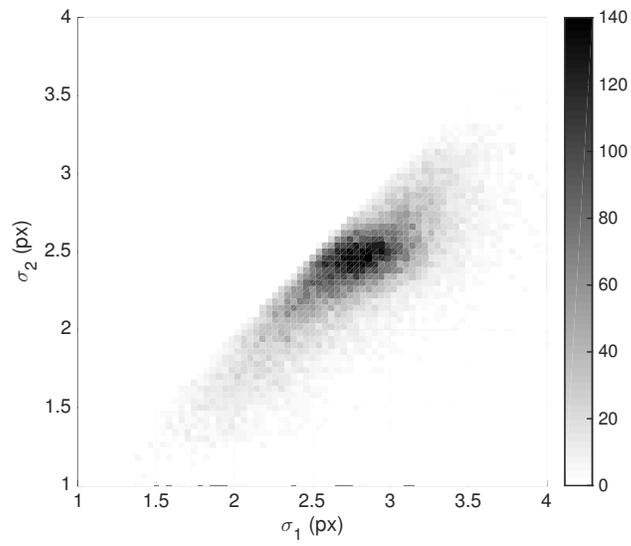} 
   \caption{2D histogram of the populations of UCN signals with respect to $\sigma_1$ and $\sigma_2$, the Gaussian widths in the major and minor semi-axes. The gray scale represents the frequency of events. Most signals populate around $\sigma_1$ of 2.76 pixels and $\sigma_2$ of 2.44 pixels. }
  \label{fig8a}
\end{figure}

\begin{figure}[htbp] 
 \centering
 \includegraphics[width=4in,angle=0]{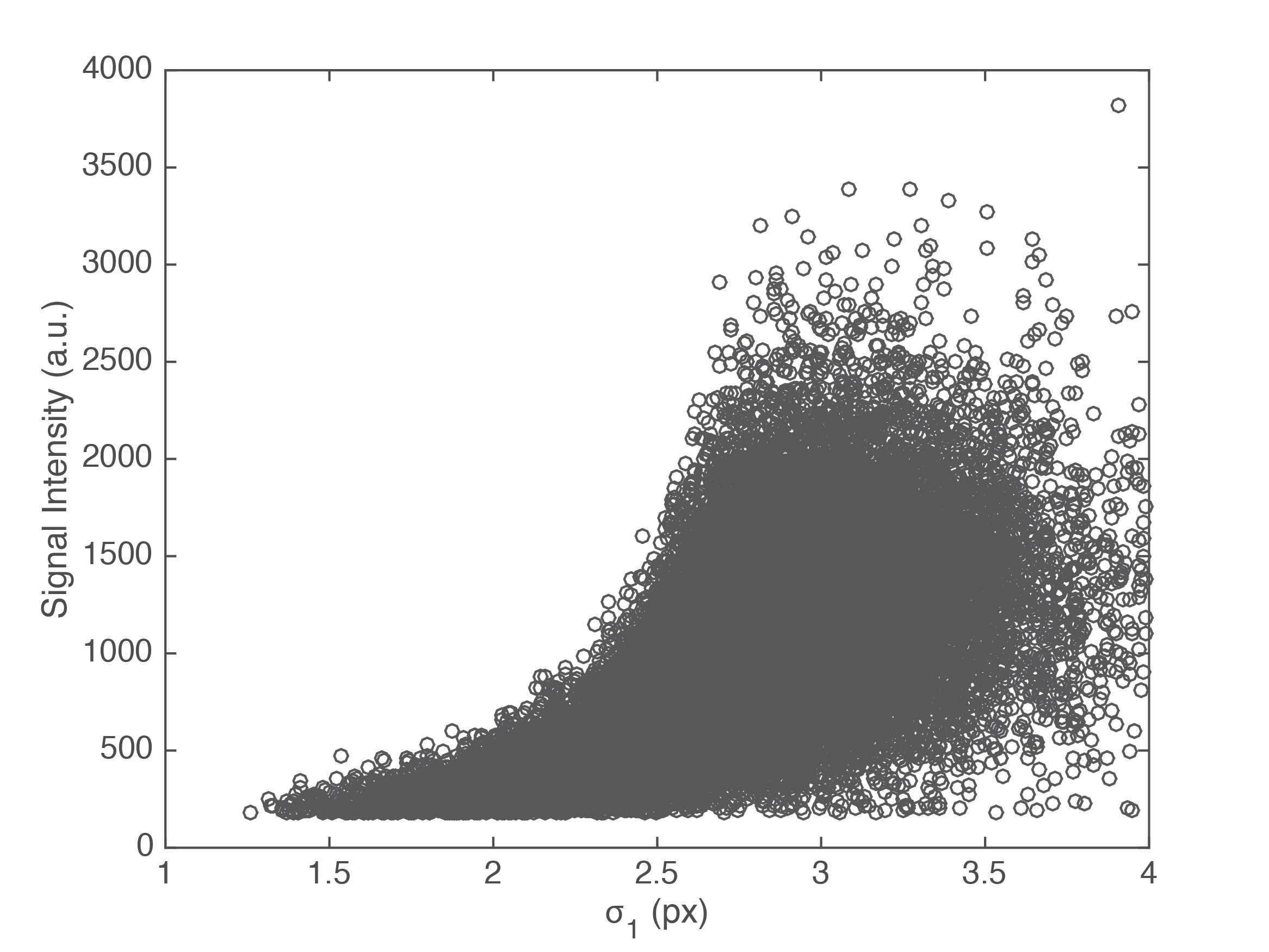}
   \caption{Signal intensity correlation with the Gaussian widths $\sigma_1$ along the major semi-axis for all the signals as shown in Figure~\ref{fig8a}. Each circle represents a scintillation signal. The most probable $\sigma_1$ of 2.77 pixels corresponds to more than 100 photons using estimates described in the text.}
  \label{fig8b}
\end{figure}

For a 2D Gaussian light intensity distribution, the error in position determination or the position resolution along one semi-axis is given by $\sigma_i/\sqrt{N_\nu}$, where $\sigma_i$ is the width of the Gaussian along the semi-axis and $N_\nu$ is the number of photons in the signal.  The fitting values of $\sigma_1$ and $\sigma_2$ for all the down-selected data are shown in Figure~\ref{fig8a}. The values of $\sigma_1$ are in the range of 1.5 to 3.8 pixels and peak around 2.77 pixels. The correlation of the peak intensities with the corresponding $\sigma_1$ value is shown in Figure~\ref{fig8b}, which indicates that $\sigma_1$ of 2.77 pixels is associated with signals of intensities around 1000. Each photon is estimated to result in an amplitude of $\sim$9 in raw camera counts, so that an amplitude of 1000 corresponds to $\sim$111 photons.
 Therefore, a typical error in position determination for a strong signal can be 0.3 pixels, or about 4 $\mu$m, along the major semi-axis. For weaker signals and larger Gaussian widths ({\it e.g.}, 20 photons and $\sigma_1$ of 4 pixels), the error of position could be as large as 0.9 pixels, which is equivalent to 13 $\mu$m. Depending on the orientation angle $\theta$, the error projecting on the Cartesian coordinate should be a propagation of errors from both semi-axes, but the maximal error does not exceed that on the major semi-axis. 

\subsection{UCN spectroscopy~\label{sec:spec}}
Accurate real-time determination of UCN positions allows precise determination of UCN energies and other kinetic information. Several methods have already been reported based on CR-39 plastic or a CCD camera\cite{Nesvizhevsky:2002,Jenke:2011,Ichikawa:2014}. Using imaging cameras has some advantages, including real-time data collection and analysis. There is no difficulty in replacing CR-39 plastic with imaging cameras to perform experiments similar to these previous works. The spatial resolution can be converted to energy resolution through the factor 102 neV/m. Therefore, a 15-$\mu$m spatial resolution is equivalent to an energy resolution of 1.5 peV, which is small enough to probe quantum bounce states of UCNs near surfaces and other fundamental physics~\cite{Jenke:2011B,Zakharov:2016}. One concern with this type of measurement is that a relatively small UCN phase space is sampled and most of the UCNs are not used. We shall discuss two types of UCN spectroscopy, one based on UCN position measurements alone, the other on both position and time-of-flight information. In both cases, larger UCN phase spaces can be sampled.

\begin{figure}[htbp] 
 \centering
 \includegraphics[width=4in,angle=0]{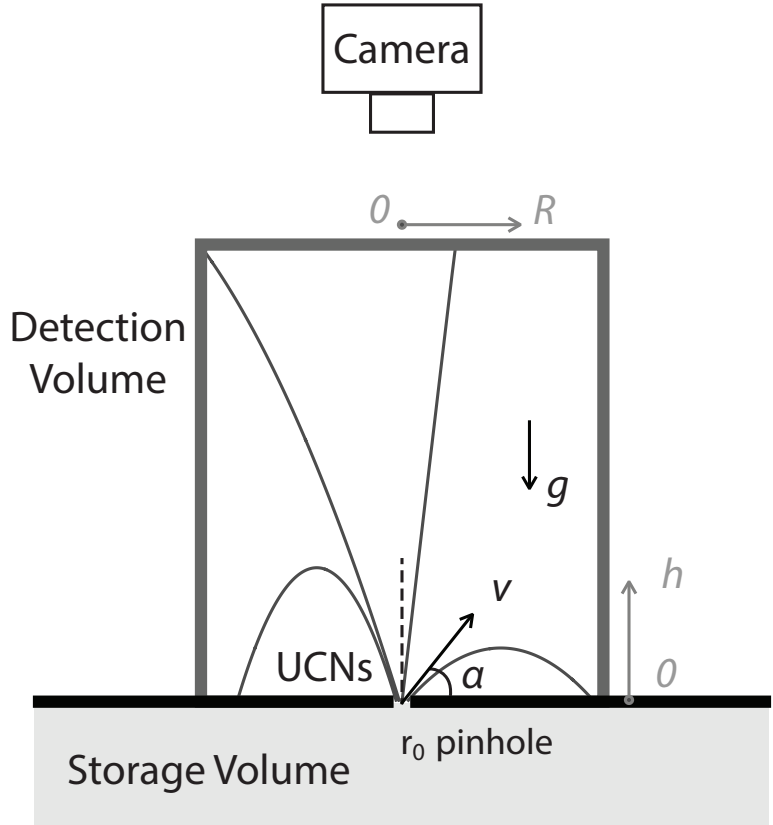} 
   \caption{The cylindrical cross section of a ToF$+$gravity hybrid UCN spectrometer. UCNs enter the detection volume via a pinhole at the ceiling of the storage volume. The detector is placed on the ceiling of the cylindrical detection volume. $\alpha$ is the angle of launch of an UCN from an entrance aperture/pinhole. Zero in height ($h=0$) is defined at the level of the pinhole.}
  \label{fig10}
\end{figure}

Measuring the energy spectrum of a UCN source is equivalent to determining the distribution function $f(E, \alpha, \phi)$, where $E$ is the UCN energy, $\alpha$ is the angle of launch as shown in Figure~\ref{fig10} and $\phi$ stands for the azimuthal angle in a cylindrical geometry. Here we assume a steady UCN source. Below, we also ignore the azimuthal variations of the distribution function. Without loss of generality, we only need to consider the 2D motion in the plane defined by the cross product of ${\bf v}\times {\bf g}$, where ${\bf v}$ is the initial velocity vector and ${\bf g}$ the vector of gravitational acceleration. The equations of motion are elementary and given by

\begin{equation}
R = R_0+ vt \cos \alpha,
\label{eq:rad}
\end{equation}
\begin{equation}
h=h_0+ vt \sin \alpha - \frac{1}{2} g t^2,
\label{eq:height}
\end{equation}
where $R_0$, $h_0 =0$ are the initial UCN coordinates or the position of the UCN entrance slit and $t$ is the time of flight. Only the magnitude of velocity $v$ is needed. We also ignore magnetic field and consider gravity as the only force during the UCN free flight.

If the time-of-flight (ToF) information $t$ is known, the initial positions $R_0$ and $h_0$, which are defined by design, and the final positions $R$, $h$, which are measured through a camera as shown in Fig.~\ref{fig10}, are sufficient to determine $v$ and the launch angle $\alpha$ uniquely through Eqs.~(\ref{eq:rad}) and (\ref{eq:height}). The measured UCN kinetic energy is given by

\begin{equation}
E = \frac{1}{2} m_0 v^2 =\frac{m_0(R-R_0)^2 + m_0 (h-h_0 + 0.5 gt^2)^2}{2 t^2},
\end{equation}
where $m_0$ is for the UCN rest mass. The energy resolution, $\delta E = m_0 v\delta v$, can be derived as

\begin{equation}
\frac{\delta E}{E} = e_1 \frac{\delta \tilde{R} }{\tilde R}+ e_2 \frac{\delta \tilde{h}}{\tilde{h}}+ e_3 \frac{\delta t}{t},
\label{eq:res}
\end{equation}
where we have used $\tilde{R} = R- R_0$ and $\tilde{h} = h -h_0$. Here $\delta t$, $\delta \tilde{R} = \sqrt{\delta R_0^2 + \delta R^2}$ and $\delta \tilde{h} = \sqrt{\delta h^2 + \delta h_0^2}$ are the measurement errors in time, the two radii and the two heights. The coefficients $e_i$'s are given by

\begin{equation}
\begin{aligned}
e_1 & = 2 \cos^2 \alpha,\\
e_2 & = \frac{\tilde{h}}{\tilde{R}} \sin 2 \alpha,\\
e_3 & =2(1 - e_1 - e_2). 
\end{aligned}
\label{eq:es1}
\end{equation}

\begin{figure}[htbp] 
 \centering
 \includegraphics[width=4in,angle=0]{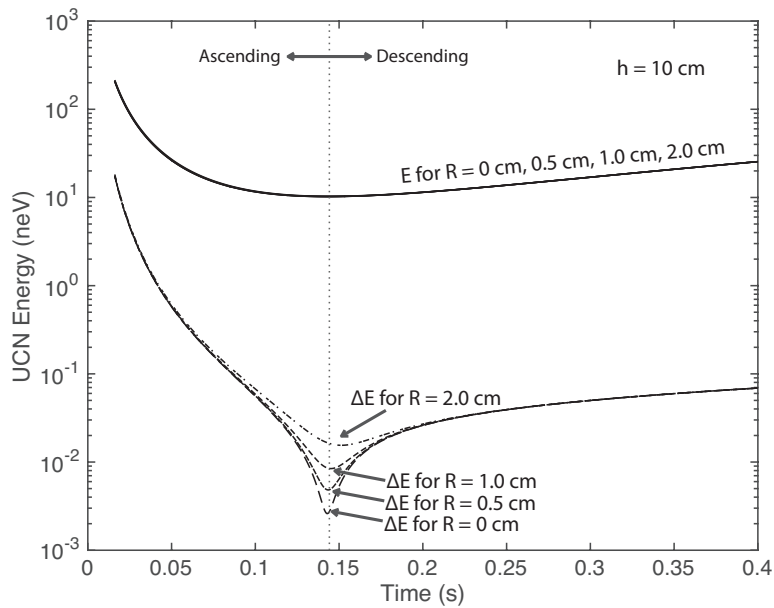} 
   \caption{Estimated UCN energy (solid lines) and associated error (dashed lines) as functions of ToF at different radius positions for a cylindrical spectrometer. The overall error is dominated by the ToF uncertainty, even for a relatively large entrance aperture of 1.6 mm in radius.}
  \label{fig11}
\end{figure}

Here we express $e_i$'s in terms of $\alpha$ for simplicity. It is straight-forward to reformulate $\alpha$ and Eqs.~(\ref{eq:res}) in terms of the measured quantities $R_0$, $R$, $h$, $h_0$ and $t$. The special case of $\tilde{R} = 0$ and $\alpha= \pi/2$ needs to be considered separately and will not be elaborated here. Based on these formulas, a spectrometer  that combines the time-of-flight with the initial and end positions can be set up. The performance of an example spectrometer of this type is shown in Figure~\ref{fig11}. The cylindrical spectrometer is 10 cm in height and 2 cm in radius with a pinhole (UCN entrance aperture) of 1.6 mm in radius. The UCN scintillation signals can be optically detected on the ceiling plate. As the height $h$ is predetermined, we assume that $\delta \tilde{h}$ comes from a typical machining accuracy of 25 $\mu$m. Here $\delta \tilde{R} $ is dominated by the pinhole radius up to 1.6 mm. The spatial resolution at the camera location (less than 15 $\mu$m) is negligible in comparison. Furthermore, if we assume the use of a fast mechanical shutter with an opening time around 1 ms at the entrance and the temporal resolution of the camera of 1 ms, the total time error $\delta t$ is 0.7 ms. The error in time $\delta t$ therefore dominates in the energy resolution in such a hybrid spectrometer. 

Figure~\ref{fig11} includes UCNs passing the height $h$ in both ascending and descending motions as if the ceiling does not exist in the detector volume. When considering the case of imaging UCN signals on the ceiling, we only need to consider the ascending section. Hence, each cycle of measurement takes about 0.15 sec. It is apparent that the UCNs of high energy arrive on the detector plane faster than the low energy ones. Within the 2 cm radius of the ceiling plate, the measured UCN energy precision is significantly affected by ToF but not the $R$ positions. The error in energy decreases from 18 neV for 200 neV UCNs to 0.32 neV for 20 neV UCNs; below 20 neV, it keeps decreasing but becomes more radius dependent; e.g., for 10.3 neV UCNs, the error in energy can achieve 16 peV at $R$ = 2 cm and 2.6 peV at $R$ = 0 cm. 

As an alternative to overcome energy resolution limited by ToF, one could choose to experimentally select the angle of launch in order to resolve the energy spectrum. Without going into details, we derive the energy and its resolution based on known angle of launch. The kinetic energy of the neutron is given by

\begin{equation}
E = \frac{m_0 g \tilde{R}^2}{4(\tilde{R} \sin \alpha \cos \alpha - \tilde{h} \cos^2 \alpha )}.
\end{equation}
The corresponding energy resolution satisfies

\begin{equation}
\delta E = f_1 \delta \tilde{R}+ f_2 \delta \tilde{h}+ f_3 \delta \alpha,
\end{equation}
where

\begin{equation}
\begin{aligned}
f_1 & = \frac{2E}{\tilde{R}}- \frac{2E^2 \sin2 \alpha}{m_0 g \tilde{R}^2},\\
f_2 & = \frac{4E^2 \cos^2\alpha}{m_0g \tilde{R}^2} ,\\
f_3 & = \frac{2E \sin \alpha}{ \cos \alpha} - \frac{4E^2}{m_0g \tilde{R}}. 
\end{aligned}
\label{eq:es2}
\end{equation}

We consider the special case when $\tilde{h} = h- h_0 = 0$, {\it i.e.}, the detectors are at the same level as the neutron entrance aperture. Similar discussions can be found, for example, in the book by Golub, Richardson and Lamoreaux~\cite{GRL:1991}. The neutron energy $E = m_0 g \tilde{R} / 2 \sin 2\alpha$. The resolution is

\begin{equation}
\frac{\delta E}{E} = \frac{\delta \tilde{R}}{\tilde{R}} - \frac{2 \cos 2 \alpha}{\sin 2 \alpha}\delta \alpha.
\end{equation}
The best energy resolution is achieved at $\alpha = \pi/4$. The energy resolution, independent of the initial neutron kinetic energy, is about 1.5 peV for $\delta \tilde{R}$ = 30 $\mu$m. From $\delta \tilde{R} = \sqrt{\delta R_0^2 + \delta R^2}$ and $\delta R =$ 15 $\mu$m at the detector location, one may use a neutron entrance as wide as $\delta R_0$ = 26 $\mu$m. For some applications when the energy resolution does not need to be this high, a wider neutron entrance aperture may be used, {\it i.e.},

\begin{equation}
\delta R_0 \sim \delta {\tilde{R}} = \frac{2\delta E}{ m_0 g}.
\end{equation}
For $\delta E = 1 $ neV, entrance aperture can be close to 2 cm wide, which is quite attractive for the UCN energy spectra measurement since a wider aperture means less time required to measure the spectra.


\subsection{Material applications}
\begin{figure}[htbp] 
 \centering
 \includegraphics[width=3in,angle=0]{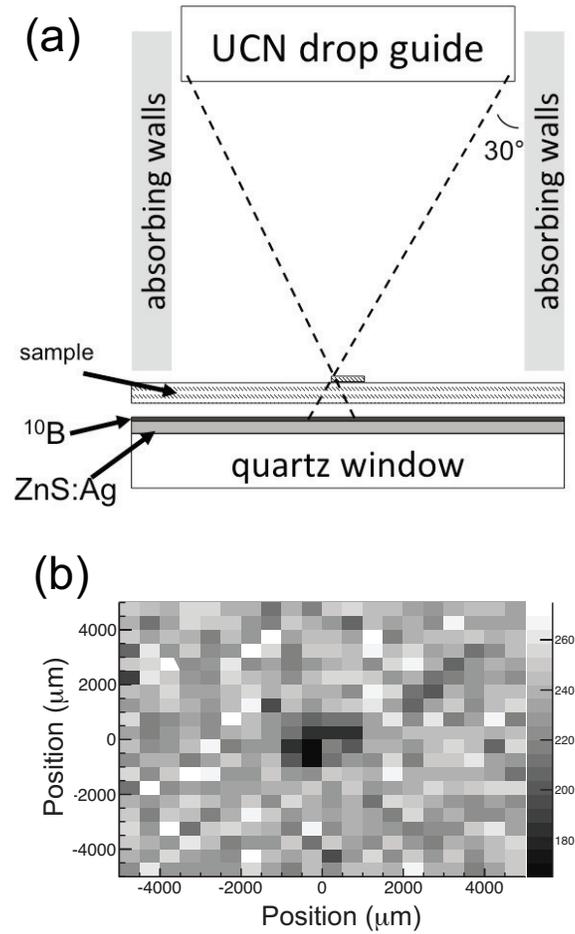} 
   \caption{Application of the position-sensitive camera to perform studies of areal variations in transmission.  (a) Example experiment geometry for transmission measurements.  UCNs drop from the guide, are collimated by absorbing walls, and pass through a thin film before being counted by the detector. (b) Monte-Carlo study of transmission through a thin film with a 1 mm diameter, 20\% absorbing spot with 10$^5$ accumulated UCN counts.}
  \label{figLJB1}
\end{figure}

Position-sensitive measurements discussed here could also enable new studies of UCN interactions with materials.  One limitation would be the very low UCN fluxes available in current sources, which preclude the use of collimation and velocity choppers to develop a monochromatic and narrow UCN ``beam".  However, rough characterizations are possible with coarse control of the distribution of neutrons and a position-sensitive detector.  UCNs could be used to characterize areal variations of thin films.  In Figure~\ref{figLJB1}a, UCNs are collimated by dropping down an absorbing guide, and a ToF chopper or a thin film with low potential could be used to select desired energies.  

The transmission as a function of position can then be measured using the detector described here.  An example of the sensitivity of such a setup was explored using a Monte-Carlo simulation.  Assuming a transmitted flux of 1 UCN/(cm$^2$ $\cdot$ s) through the sample and maximum incident angle of 30$^o$, and neglecting reflection (UCN energy E $>$ material Fermi potential), a 1 mm diameter spot with 20\% lower transmission (such as due to surface contamination or variations in thickness) on a 1 mm thick sample could be resolved after about 1 day of measurement, Figure~\ref{figLJB1}b. With control of the range of UCN energies from the loss due to reflection from the material potential, variations in density or isotopic concentrations could also be studied.


\section{Conclusion}

We have successfully demonstrated position-sensitive measurements of UCNs by optically coupling a multilayer surface detector to an imaging CCD camera with 16-$\mu$m pixels. The UCN position accuracy is as small as 4 $\mu$m for a typical strong scintillation signal and about 13 $\mu$m for weak ones. Several feasible UCN spectroscopy concepts are discussed with the best energy resolution below 2 peV. Therefore, this method can be employed as a real-time detector in studying UCN quantum states in the gravitational field. 
Other possible applications include UCN microscopy and reflectometry for material science.

{\bf Acknowledgments} We gratefully acknowledge the support of the U.S. Department of Energy through the LANL/LDRD Program. L.J.B. is partly supported by the G. T. Seaborg Institute and Los Alamos Science Campaign C1 for this work.

\bibliographystyle{elsarticle-num}
\bibliography{UCNImagingv2}







\end{document}